\documentclass[useAMS,usenatbib,letterpaper,10pt]{mn2e}
\usepackage{times}
\usepackage{epsfig}
\usepackage{natbib}
\usepackage{amssymb}
\usepackage{amsmath}
\usepackage[bookmarks,bookmarksnumbered,colorlinks=true, citecolor=blue, linkcolor=black]{hyperref}
\usepackage{color}
\usepackage[normalem]{ulem}

\newcommand{\Msun}{M$_\odot$} 
\defcitealias{Torrey2015}{T15}
\defcitealias{Torrey2016}{T16}

\title[Predicting progenitor properties probabilistically]{An improved probabilistic approach for linking progenitor and descendant galaxy populations using comoving number density}
\author[Wellons \& Torrey]{Sarah Wellons$^1$\footnotemark[1] and Paul Torrey$^{2,3}$ \\
$^1$Harvard-Smithsonian Center for Astrophysics, 60 Garden St., Cambridge, MA 02138, USA \\
$^2 $ MIT Kavli Institute for Astrophysics \& Space Research, Cambridge, MA, 02139, USA\\
$^3 $ TAPIR, Mailcode 350-17, California Institute of Technology, Pasadena, CA 91125, USA }
\begin{document}

\maketitle

\label{firstpage}

\begin{abstract}
Galaxy populations at different cosmic epochs are often linked together by comoving cumulative number density in observational studies.  Many theoretical works, however, have shown that the number densities of tracked galaxy populations evolve in bulk and spread out over time.  We present a number density method for linking progenitor and descendant galaxy populations which takes both of these effects into account.  
We define probability distribution functions that capture the evolution and dispersion of galaxy populations in comoving number density space, and use these functions to assign galaxies at one redshift $z_f$ probabilities of being progenitors or descendants of a galaxy population at another redshift $z_0$. 
These probabilities are then used as weights for calculating distributions of physical properties such as stellar mass, star formation rate, or velocity dispersion within the progenitor/descendant population.  
We demonstrate that this probabilistic method provides more accurate predictions for the evolution of physical properties then either the assumption of a constant number density or the assumption of an evolving number density in a bin of fixed width by comparing the predictions against galaxy populations directly tracked through a cosmological simulation.
We find that the constant number density method performs most poorly at recovering galaxy properties, the evolving number method density slightly better, and the probabilistic number density method best of all.  
The improvement is present for predictions of both stellar mass as well as inferred quantities such as star formation rate and velocity dispersion which were not included in the number density fits.
We demonstrate that this method can also be applied robustly and easily to observational data, and provide a code package for doing so.
\end{abstract}


\renewcommand{\thefootnote}{\fnsymbol{footnote}}
\footnotetext[1]{E-mail: swellons@cfa.harvard.edu}

\section{Introduction}

Extragalactic observations have provided us with a wealth of data about how the global population of galaxies grows and evolves throughout cosmic time.  We have learned about the rise and fall of the cosmic rate of star formation, the buildup of stellar mass in galaxies, and the emergence of a population of a massive, quiescent galaxies.  A challenge which still remains, however, is to describe how individual galaxies (or galaxy populations) evolve within this global population, and to draw connecting lines between galaxies at different redshifts.  

One method which is commonly used to make these connections is the assumption of a constant cumulative comoving number density \citep{VanDokkum2010}.  In this method, galaxies are ranked according to their stellar mass and assigned the comoving number density of all galaxies which are at least as massive (this is known as the ``cumulative mass function" or CMF).  Under the assumptions that (a) the number of mergers is negligible, and (b) the rank order of galaxies is preserved across time, the comoving cumukative number density of each galaxy will remain constant since there will always be the same number of galaxies which are more massive than they.  Thus, armed solely with a set of CMFs, one can predict the progenitors or descendants of a population of galaxies by identifying its number density\footnote{For brevity, we refer to ``cumulative comoving number density" simply as ``number density" hereafter.  All number densities should be assumed to be cumulative and comoving unless explicity marked otherwise.} at one redshift and finding the stellar mass corresponding to that same number density at other redshifts.  This method has been used to predict the evolution of galaxies' stellar masses \citep{VanDokkum2013, Brammer2011, Muzzin2013} as well as numerous other properties including star formation rate \citep{Papovich2011, Fumagalli2012}, velocity dispersion \citep{Bezanson2011}, gas content \citep{Conselice2013}, and size \citep{Patel2013}.

Theoretical studies of galaxy formation, on the other hand, do allow us to follow the evolution of individual galaxies throughout time.  Many of these studies have shown that neither of the aforementioned assumptions for constant number density are strictly true: mergers do remove galaxies from the population (even at the massive end), and rank order is not preserved, resulting in the evolution of galaxies' number densities with time.  This has been observed in abundance matching models \citep{Behroozi2013}, semi-analytical modeling \citep{Leja2013}, and hydrodynamical simulations at high redshift \citep{Jaacks2016} through the present day \citep[hereafter T15 and T16]{Clauwens2016, Torrey2015, Torrey2016}.  These various theoretical methods have painted a surprisingly consistent picture of how galaxy number densities evolve.  In \citetalias{Torrey2015}, we examined how galaxy populations in the cosmological hydrodynamical simulation Illustris evolve in number density from $0 < z < 3$, and found fits to the number density evolution which are strikingly similar to those from \citet{Behroozi2013} and \citet{Leja2013}.  In addition to the consistency between theoretical methodologies, we also found that the number density evolution was equivalent regardless of whether stellar mass, dark matter halo mass, or stellar velocity dispersion was used to assign galaxy rank.  

This consistency points to a fundamental stoschasticity driving the number density evolution which stems from the $\Lambda$CDM framework underlying all these works.  In the case where galaxy stellar mass is used to assign rank, this manifests as stochasticity in the galaxy merger rates and star formation rates driven by variations in the accretion rate of gas into halos.  If the number density evolution is driven by $\Lambda$CDM, then we expect that it should be robustly applicable to observational datasets.  The median number density evolution provided by \citet{Behroozi2013} has in fact already been incorporated into observational studies of local ultra-massive galaxies \citep{Marchesini2014}, the star formation histories of high-redshift galaxies \citep{Salmon2015}, and the progenitors of M* galaxies \citep{Papovich2015}.

However, the median evolution in number density does not capture the full behavior of galaxy populations.  In addition to a net evolution, the number densities of galaxies also {\it spread out} over time, even if they were tightly grouped together originally.  Some examples may be seen in \citet{Wellons2015, Wellons2016}, wherein we followed a group of $\sim10^{11}$ \Msun~compact galaxies at $z=2$ from the Illustris simulation back in time to their high-redshift progenitors and forward in time to their $z=0$ descendants.  In both directions, the galaxies undertook a variety of evolutionary paths and their stellar masses spread out enormously (covering an order of magnitude at $z=0$).  
\citet{Terrazas2016} have recently shown a similarly diverse set of progenitors for Milky Way-like galaxies in semi-analytical models.  To properly characterize the evolution of these populations, simply quoting a median mass is insufficient -- we also need to consider the width of the stellar mass (or number density) distribution.  For the majority of the progenitors/descendants, the median evolution would be a poor representation.  

To improve predictions of the progenitors and descendants of galaxy populations, we need to acknowlege that galaxies do not evolve in lockstep.  When discussing their evolution, we should use probabilistic rather than deterministic language, consider entire distributions instead of medians, and eschew the use of the word ``typical" in reference to galaxy progenitors and descendants.  In this paper, we present a modified number density method for predicting the properties of progenitor and descendant populations.  This method takes into account not only the median evolution in number density, but also the width of the probability distribution around which it is centered.  We will show that it produces more accurate predictions in simulations where we know the true progenitors, and provide code that can be robustly and easily applied to observational datasets.\footnote{\url{https://github.com/sawellons/NDpredict}}

\begin{figure*}
  \centering
  \includegraphics[width=2.1\columnwidth]{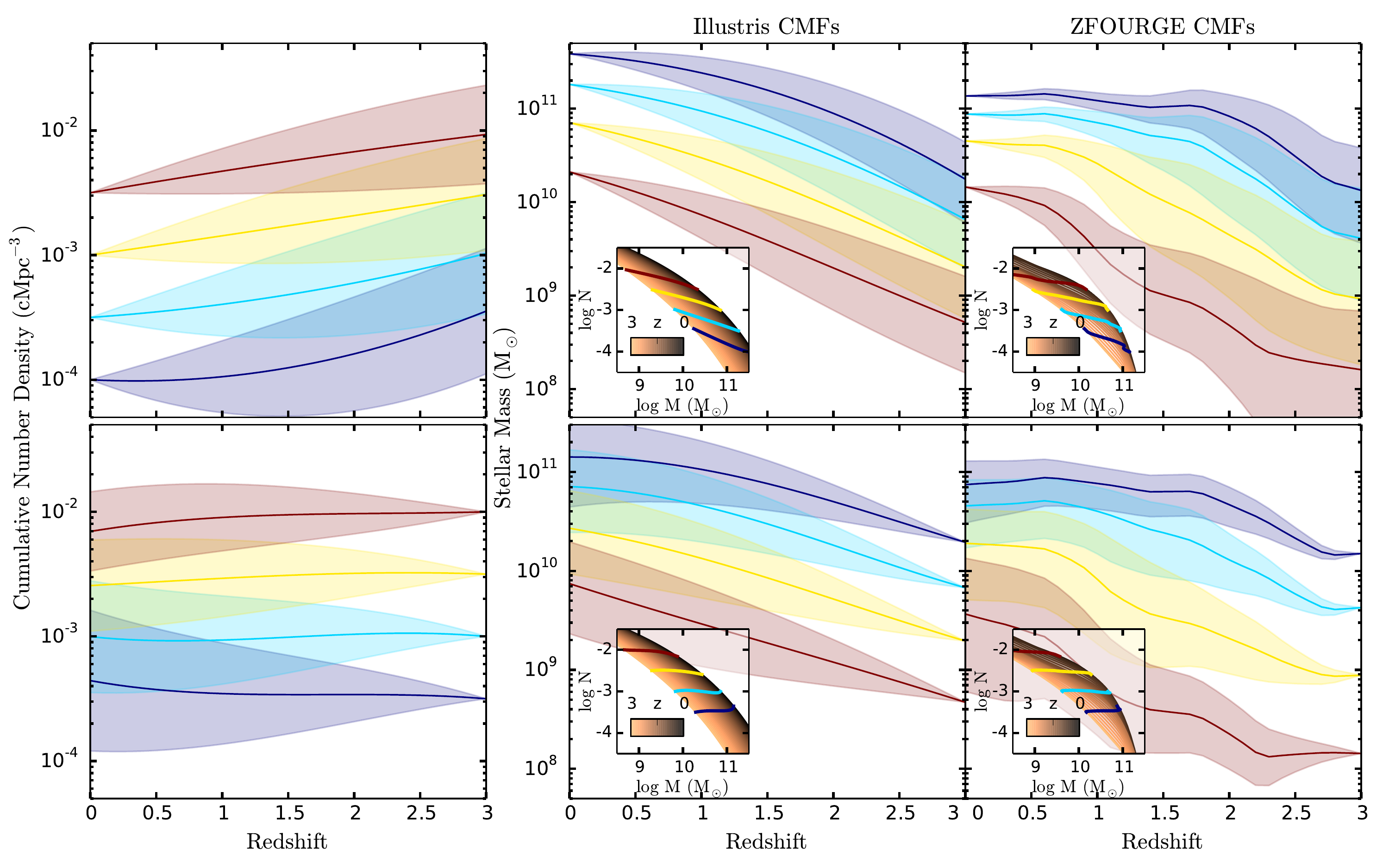}
  \caption{Evolution in number density and stellar mass of galaxy populations between $z=0$ and $z=3$.  In the top row, populations are selected by number density at $z=0$ and traced backwards in time, and in the bottom row galaxy populations are selected at $z=3$ and traced forward.  {\it Left:} Number density evolution.  Solid lines show the fit from \citetalias{Torrey2016} to the median number density of the population and shaded regions the 1$\sigma$ contours, as described in Section \ref{ssec:Nfits}.  {\it Middle:}  Stellar mass evolution corresponding to the number density evolution on the left, when the CMFs from the Illustris simulation are used to convert between number density and stellar mass.  The CMFs from $z=3$-0 appear in an inset in the bottom left corner, and the trajectories traced out by the median of each population are shown by lines of the corresponding color.  {\it Right:} Same as the middle column, except that the observational CMFs from ZFOURGE are used to assign stellar mass.}
  \label{fig:Nevolfits}
\end{figure*}

The structure of the paper is as follows: In Section \ref{sec:methods}, we describe our probabilistic number density method for predicting galaxy progenitor/descendant populations.  In Section \ref{sec:MW}, we use the progenitors of Milky-Way-mass galaxies as a test case to study the predictions made by various number density methods, including the assumption of a constant number density, an evolving number density, and a probabilistic number density.  We apply these methods both to data from the Illustris hydrodynamical simulation as well as to an ``observational" dataset which follows observed relations.  In Section \ref{sec:discussion}, we discuss the overall accuracy of these predictive methods as well as their limitations and possible refinements.  We conclude in Section \ref{sec:conclusion}.

\section{Methods}
\label{sec:methods}

\subsection{Number density evolution}
\label{ssec:Nfits}

In \citetalias{Torrey2016}, we defined prescriptions for the number density evolution of galaxy populations tracked through the cosmological hydrodynamical simulation Illustris \citep{Vogelsberger2014, Vogelsberger2014a, Genel2014, Nelson2015}.  We argued that the number densities of the descendants of galaxy populations with initial number density $N_0$ at redshift $z_0$ spread into approximately lognormal distributions at another redshift $z_f$ which can be characterized by a median $\hat{N}_f$ and width $\sigma_{\log N}$.  We tabulated fits for $\hat{N}_f$ and $\sigma_{\log N}$ as a function of $N_0, z_0$, and $z_f$ which can be used to plot the evolution of galaxies in number density space.  

Example tracks are shown in the left column of Figure \ref{fig:Nevolfits}, where solid lines indicate the median\footnote{$\hat{N}$ is related to $\mathcal{N}$ from \citetalias{Torrey2016} as $\langle \mathcal{N} \rangle = \log \hat{N}$.} $\hat{N}$ and colored regions the intrinsic $1\sigma_{\log N}$ scatter for galaxy populations traced between $z=0$ and $z=3$.  Note that if the constant number density assumption were satisfied, the lines would appear horizonal.  The top row shows populations selected in number density at $z=0$ and traced backwards in time, while the bottom row shows galaxy populations selected at $z=3$ and traced forward in time.  The number density evolution tracks described in \citetalias{Torrey2016} incorporate both the impact of scattered growth rates and galaxy coagulation (i.e. mergers).  For further details, including the relative importance of scattered growth rates and coagulation and an exploration of progenitor and descendant tracking asymmetry, see \citetalias{Torrey2016}.

In this paper, we consider the impact of applying these number density evolution tracks to observational data.  In the following analysis, we will primarily examine progenitor populations (traced backward in time) as examples, but everything which follows is equally applicable to descendant populations (traced forward in time) as well.

\subsection{Mass functions}
\label{ssec:massfuncs}

The number density evolution tracks found in cosmological hydrodynamical simulations, semi-analytic models, and abundance matching predictions are  similar.  As was argued in \citetalias{Torrey2015}, this is caused by the number density evolution tracks being a robust prediction of $\Lambda$CDM based on the underlying growth of the dark matter halos.  Converting this robust number density evolution to the evolution of a physical quantity (e.g. stellar mass), however, requires a cumulative mass function (CMF) which ranks galaxies according to that quantity.  In this paper, we consider the evolution of galaxy properties using number density analyses based on two different sets of galaxy stellar mass functions from the Illustris simulation and the ZFOURGE observational dataset \citep{Tomczak2014}.

For comparisons with simulation results, we adopt the Illustris CMF tabulated in \citetalias{Torrey2015}.  This CMF is provided as a single function that is valid over the mass range $10^7$ \Msun~$< M_* <10^{12}$ \Msun, redshift range $0<z<6$, and number density range $\phi > 3 \times 10^{-5}$ Mpc$^{-3}$ dex$^{-1}$.  Employing the Illustris CMF facilitates a comparison between the mass, SFR, and velocity dispersion evolution predicted by the number density methods against the directly calculated mass, SFR, and velocity dispersion evolution derived using the simulation merger trees.

When making observational predictions, we adopt the ZFOURGE mass functions using the tabulated double-Schechter functions provided in \citet{Tomczak2014}.  These fits are valid over a mass range of $10^8 < M_*/$\Msun $< 10^{11.5}$ at redshifts $0.2 < z < 3$.  The CMF in each redshift bin is computed by integrating the fits from $M_*$ to infinity using the mpmath package \citep{Johansson2014}.  We find the CMF at an arbitrary redshift by linearly interpolating in $\log N$ and $z$ between the CMFs in the adjacent redshift bins.

The choice of mass function is important because it transforms the evolution from number density space to physical space.  A single evolutionary track in number density can manifest as different evolutionary tracks in stellar mass depending on the mass function employed.  This concept is illustrated in Figure \ref{fig:Nevolfits}, which shows the stellar mass evolution inferred using the Illustris (middle column) and ZFOURGE (right column) CMFs from the same number density evolution (left column).  The CMFs themselves appear in an inset in the bottom left of each panel and range from $0 < z < 3$.  Colored lines moving across the CMFs trace out the trajectory of the corresponding lines in the larger panels.  The shape of the chosen CMF has a direct effect on the predicted mass evolution.  For example, the ZFOURGE mass functions are steeper at the high-mass end, which drives a shallower stellar mass evolution for massive galaxies.  This figure also demonstrates that the number density evolution is equally applicable to data from simulations and observations, and that although the number density evolution fits were made using simulation data, the simulation's mass function is not required in order to use them.

\subsection{Predicting progenitor populations}
\label{ssec:predictions}

Throughout this paper, we will explore three methods for selecting progenitor galaxy populations based on number density.  The following subsections describe these methods for predicting the progenitors at redshift $z_f$ of a population of galaxies which were selected to have an initial number density $N_0 \pm dN$ at redshift $z_0$:

\subsubsection{Constant number density}

Galaxies are selected in the same, constant number density range $N_0 \pm dN$ at $z_f$ as their descendants.  The galaxy population thus selected will have a new stellar mass according to the appropriate CMF at that redshift.  This is the original implementation of constant comoving number density \citep{VanDokkum2010} as applied in e.g. \citet{Papovich2011, Bezanson2011, Muzzin2013} and \citet{VanDokkum2013}.

\subsubsection{Evolving number density}

The new median number density of the population $\hat{N}_f(N_0, z_0, z_f)$ is calculated using Equation A4 (for progenitors, $z_f > z_0$) or A1 (for descendants, $z_f < z_0$) of \citetalias{Torrey2016}.  Galaxies are selected in the number density range $\hat{N}_f \pm dN$ according to the CMF at that redshift.  This is similar the the modified, non-constant comoving number density applied in e.g. \citet{Marchesini2014, Salmon2015}, and \citet{Papovich2015} using functions from \citet{Behroozi2013}.

\begin{figure*}
  \centering
  \includegraphics[width=2\columnwidth]{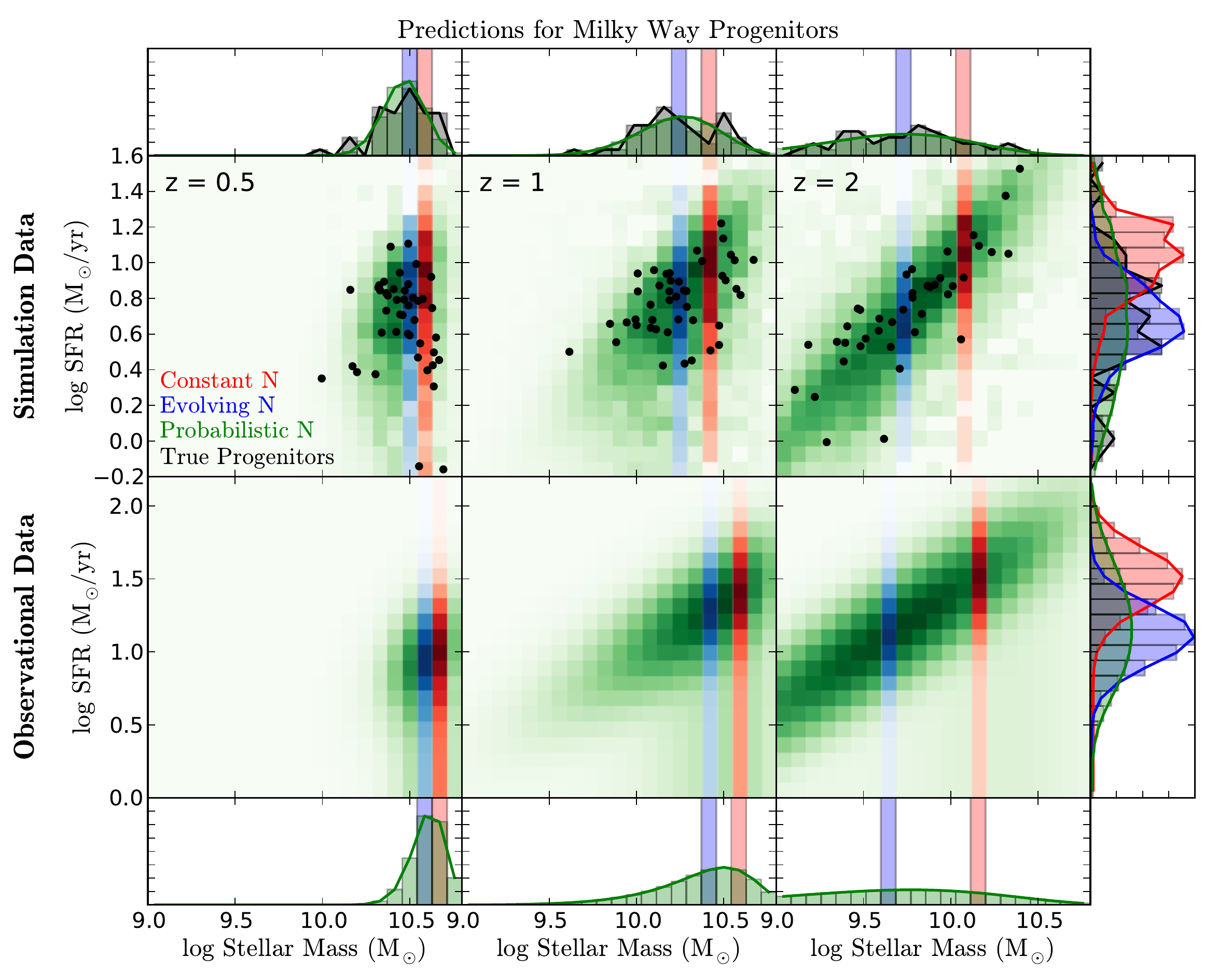}
  \caption{Predictions for the $z= [0.5, 1, 2]$ stellar mass and star formation rates of progenitors of galaxies that had stellar mass $10^{10.6}$ \Msun~at $z=0$.  Central panels show 2d histograms of the probability density predicted using a constant number density (red), evolving number density (blue), and probabilistic number density (green).  Top panels show predictions using data from the Illustris simulation and include black points for the true progenitors.  Bottom panels show predictions using a sample constructed from observational relations (see Section \ref{ssec:sample}).  One-dimensional histograms to the top and bottom show the projected distribution of stellar masses in the adjacent panel, and those to the right show the projected distribution in star formation rate at $z=2$.  Each method predicts a different distribution in stellar mass, which in turn produces a different prediction for star formation rate.}
  \label{fig:2dpredictions}
\end{figure*}

\subsubsection{Probabilistic number density}
\label{sssec:distrib}

Calculate the new median number density $\hat{N}_f(N_0, z_0, z_f)$ and distribution width $\sigma_{\log N}(N_0, z_0, z_f)$ using Equations A4 and A5 (for $z_f > z_0$) or A1 and A2 (for $z_f < z_0$) from \citetalias{Torrey2016}.  These quantities define a lognormal probability distribution function 
\begin{equation} \frac{dp}{d\log N_f} = \frac{1}{\sqrt{2 \pi \sigma_{\log N}^2}} \exp \left( - \frac{(\log N_f - \log \hat{N}_f)^2}{2 \sigma_{\log N}^2} \right) \end{equation}
which is itself determined by $N_0$, $z_0$, and $z_f$.  This pdf $dp/d\log N_f$ describes the probability of a progenitor galaxy at $z_f$ having number density $N_f$.  

The number density probability distribution can then be used to assign {\it every} galaxy in the sample at $z_f$ a probability of being a progenitor $p_{\rm prog}$ according to its number density.  (From here on we will speak only in terms of looking back toward progenitors, but there is an equivalent $p_{\rm desc}$ when looking forward toward descendants.)  For a galaxy with number density $N$, its progenitor probability is
\begin{equation} 
p_{\rm prog} = \frac{1}{V} ~ \frac{1}{N} ~ \left. \frac{dp}{d\log N_f}\right|_N
\end{equation}
where the first factor of $1/V$ is the discrete step in number density for a single galaxy in the sample (for the Illustris sample, this is 1/(106.5 Mpc)$^3$) and the second factor of $1/N$ converts between logarithmic and linear number densities.  This $p_{\rm prog}$ describes the likelihood that a given galaxy at $z_f$ is a progenitor of a galaxy in the original $z_0$ population.  Note that the sum over the $p_{\rm prog}$s of every galaxy at $z_f$ is equal to 1, which can be understood to mean that you are guaranteed to find the progenitor of any given galaxy if you search over the entire sample.

Having calculated the progenitor probability for every galaxy in the sample, the distributions of physical properties (e.g. stellar mass, star formation rate, velocity dispersion, size, etc.) in the progenitor population can be predicted by using the progenitor probabilities as weights and summing over the entire sample.  As a concrete example, the mean stellar mass of the progenitors may be determined via
\begin{equation} \langle M_* \rangle = \sum_i^{\rm all~galaxies} p_{{\rm prog}, i}~M_{*,i} \end{equation}
where $p_{{\rm prog}, i}$ and $M_{*,i}$ are the progenitor probabilities and stellar masses of individual galaxies in the sample.
Similarly, the probability that a progenitor galaxy would have a star formation rate exceeding 100 \Msun/yr is the sum over the progenitor probabilities of all galaxies which meet that criterion.
In practice, in both cases one will also have to divide out $\sum_i^{\rm all} p_{{\rm prog}, i}$ unless the entire relevant number density range is well-sampled within the dataset.

The key distinction between this method and those above is that we are not fixing the progenitors/descendants to a single number density (stellar mass).  Instead, this method encompasses the full range of evolutionary paths that galaxies might undertake, and affords us the flexibility to speak both in terms of full distributions as well as means and medians.

\subsection{Galaxy samples}
\label{ssec:sample}

The prescriptions described in the previous subsection provide methods for selecting progenitor/descendant galaxies from a dataset based on number density as assigned by stellar mass.  In later sections of this paper, we will show the predictions generated by these methods using two different galaxy samples: a simulation sample, and an observational sample.  The simulation sample allows us to make comparisons between predicted progenitors and true progenitors.  The observational sample demonstrates the applicability of all three methods to observational data, and gives a sense of how observational results might be affected by the choice of method.

\subsubsection{Simulation sample}

The Illustris simulation \citep{Vogelsberger2014, Vogelsberger2014a, Genel2014, Nelson2015} has a comoving volume of (106.5 Mpc)$^3$ and contains tens of thousands of galaxies, with well-resolved and well-sampled galaxy populations in the $M_* \approx 10^{9-12}$ \Msun~range.  When selecting progenitors from this sample, we choose them directly out of the simulation volume at the appropriate redshift.  The number densities of these simulated galaxies are assigned according to the Illustris CMFs discussed above.  The crucial advantage to using simulation data is that the galaxy samples at different redshifts are direct progenitors/descendants of one another.  Individual galaxies are connected via merger trees \citep{Rodriguez-Gomez2015} which enable us to identify the ``true" progenitors for comparison with the predicted progenitors.

\subsubsection{Observational sample}

In practice, when using the number density methods described in this paper one would be drawing from an observational sample in much the same way as we do with the simulation sample above.  To get a broad sense of what one might expect when doing so, we construct a mock observational dataset generated from observed relations between galaxy properties.  The sample thus produced contains galaxy stellar masses, star formation rates, effective radii, and velocity dispersions which follow the observed quenched fraction, star formation main sequence, size-mass relations, and fundamental plane at all masses and redshifts.

In detail, when ``selecting" a galaxy with number density $N$ we first convert that number density to stellar mass with the ZFOURGE mass functions as described above.  \citet{Tomczak2014} also break the mass function down into star-forming and quiescent components, from which we can infer $f_{\rm quench}(M_*)$, the probability of a galaxy of a given stellar mass being quenched.  We decide whether the galaxy is star-forming or quiescent probabilistically based on $f_{\rm quench}$.  If it is star-forming, we select a star formation rate (SFR) from a log-normal distribution with a width of 0.2 dex centered around the main-sequence value taken from \citet{Speagle2014}.  If it is quiescent, we draw a SFR from a uniform distribution in log space from 0.6-3 dex below the main-sequence value.  We also use the stellar mass, redshift, and quiescent/star-forming status to define an appropriate probability distribution for effective radius from \citet{VanderWel2014} and draw a value.  Finally, we use the stellar mass, effective radius, and redshift to determine the galaxy's stellar velocity dispersion from the fundamental plane fit from \citet{Bezanson2013}.  

We emphasize that the ``predictions" we make with this sample are intended simply to demonstrate the sort of changes one might expect to see with different number density methods when using observational datasets.  The sample is designed to be reasonably similar to observations, but is destined to be incorrect in detail.  The distribution of star formation rates for quenched galaxies, for example, is poorly known, and to reach the low stellar masses we discuss in the next section we must extrapolate the fundamental plane fit to regions where it is not constrained.  Thus the exact evolution we show here should not be considered a true prediction, but rather a proof-of-concept.

\section{A test case: Milky Way progenitors}
\label{sec:MW}

\begin{figure*}
  \centering
  \includegraphics[width=2.1\columnwidth]{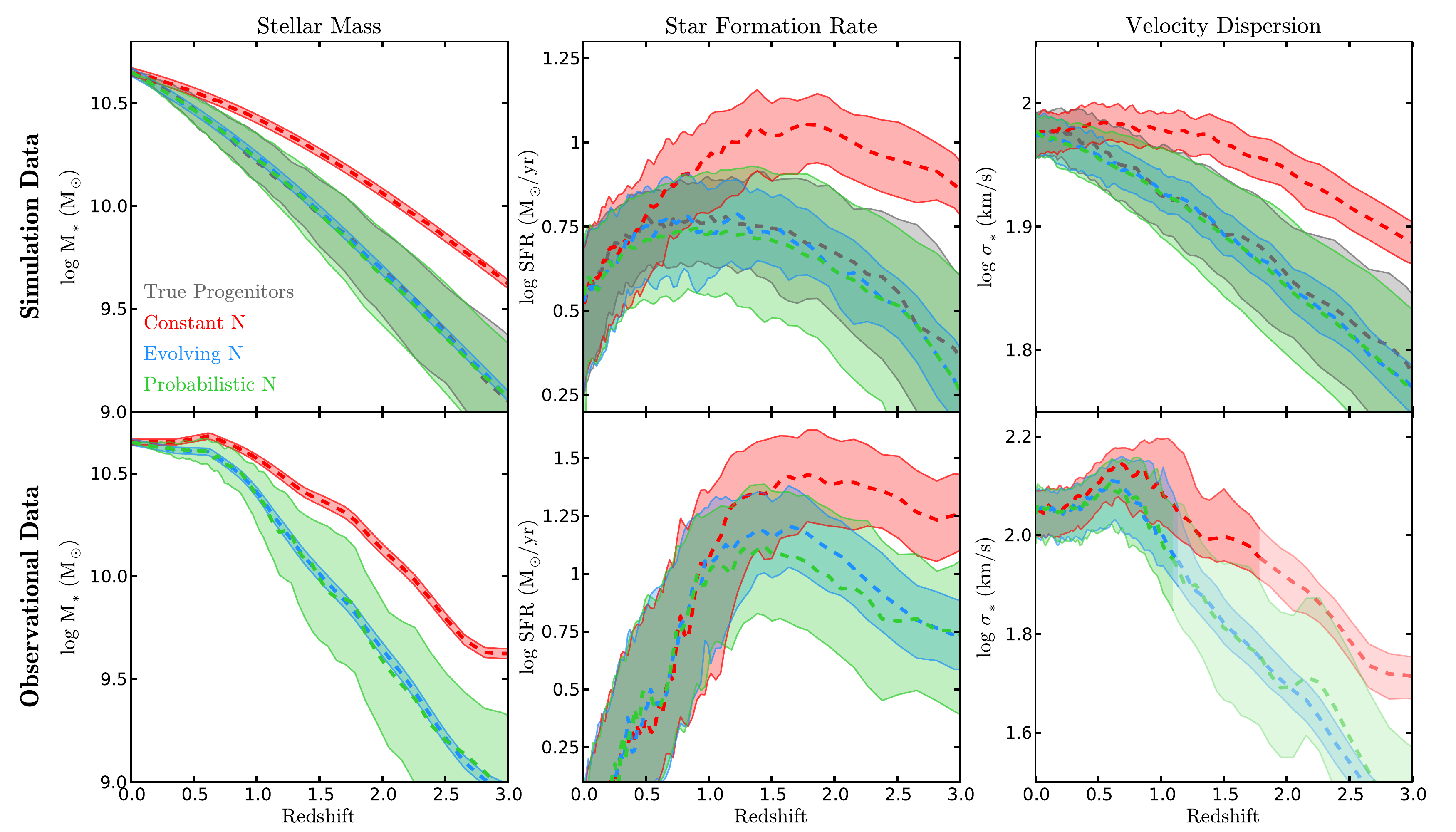}
  \caption{Predicted evolution of the stellar mass, star formation rate, and stellar velocity dispersion of the progenitors of a population of 300 galaxies with stellar mass $10^{10.6}$ \Msun~at $z=0$, traced back to $z=3$.  The dashed lines trace the median values within the predicted populations, and shaded regions the 30-70th percentile region.  Line color indicates the predictive method employed: red for a constant number density, blue for an evolving number density, and green for a probabilistic number density.  The top row shows predictions using galaxies from the Illustris simulation, and also includes grey lines for the true progenitors traced through the simulation.  The bottom row shows predictions derived from observational relations (see Section \ref{ssec:sample}).  We fade out the observational velocity dispersion lines below log $\sigma_*$ = 1.95, where the fit to the fundamental plane is an extrapolation.  In general, the evolving and probabilistic methods accurately predict the median evolution of these quantities, and the probabilistic method also captures the width of the population distribution.}
  \label{fig:predictfromsample}
\end{figure*}

In this Section we explore the predictions made by each of the progenitor/descendant linking methods for Milky Way (MW) progenitors.  
Our $z=0$ sample of MW-mass galaxies is chosen such that their stellar masses are centered around $10^{10.6}$ \Msun.  We convert this initial mass to an initial number density $N_0$ using the $z=0$ CMF, and select $n=50$ galaxies from the number density bin $N_0 - n/2V < N < N_0 + n/2V$.  

We then predict how the progenitors will evolve according to each of the three number density methods described in Section \ref{ssec:predictions}.  All predictions are made twice: once using the Illustris CMFs and drawing progenitor galaxies from the simulation, and once using the ZFOURGE CMFs and drawing progenitor galaxies from the mock observational sample described in Section \ref{ssec:sample}.  In the Illustris case, we also follow the galaxies in the initial sample back to their true progenitors for comparison with the predictions.

Figure \ref{fig:2dpredictions} shows the two-dimensional probability distribution for 50 MW progenitors in stellar mass and star formation rate at $z$ = 0.5, 1 and 2.  The constant and evolving number density predictions appear as vertical strips of red and blue respectively because these methods necessarily predict a very narrow range in stellar mass.  The probabilistic prediction is shown in green and covers a wide range of stellar masses.  In this case, the histogram includes every galaxy in the full sample at that redshift, weighted by its progenitor probability assigned as described in Section \ref{sssec:distrib}.  The black points in the top row represent the true progenitors of the galaxies in the original $z=0$ sample from the Illustris simulation.  One-dimensional histograms to the top, bottom, and right show the projected distributions in stellar mass and star formation rate separately.  

The evolving number density prediction more accurately captures the median stellar mass of the true progenitors when compared against the constant number density method, as can be seen in the top row of panels.  Despite capturing this median evolution, the evolving number density method fails to reproduce the full distribution of progenitor stellar masses (compare the black and blue histograms).  To recover this distribution, the probabilistic method (green histogram) is required. 

The predicted progenitor stellar mass distribution impacts secondary inferred quantities, such as the star formation rate.  Since the star formation rate was not considered when determining the number density evolution fits, the black and green histograms for SFR (on the right of Figure \ref{fig:2dpredictions} do not overlap as neatly as the ones for stellar mass.  However, the inferred SFR distribution for the probabilistic linking method still provides an improved match to the true progenitor distribution when compared against the constant and evolving number density selections.  The improvement in reproducing progenitor galaxy properties is quantified in Section \ref{ssec:accuracy}.

The bottom row of the figure shows the same predictions using the observational sample.  In this case, the true progenitor properties are not known.  However, it is again clear that the different progenitor/descendant linking methods produce qualitatively different predictions. 

In Figure \ref{fig:predictfromsample}, we show the evolution of a population of 300 MW progenitors as a continuous function of redshift.  The three methods are shown using the same color scheme as the previous figure. Dashed lines indicate the median stellar mass, star formation rate, and stellar velocity dispersion of the predicted progenitor population.  The shaded region indicates the 30th-70th percentile range.  As in Figure \ref{fig:2dpredictions}, the values for the probabilistic method (in green) are computed using the progenitor probabilities as weights.  Results from the Illustris simulation appear in the top row, and results from the observational sample in the bottom row.  The grey line and regions in the Illustris panels refer to the true progenitor population traced through the simulation.

\begin{figure*}
  \centering
  \includegraphics[width=2.1\columnwidth]{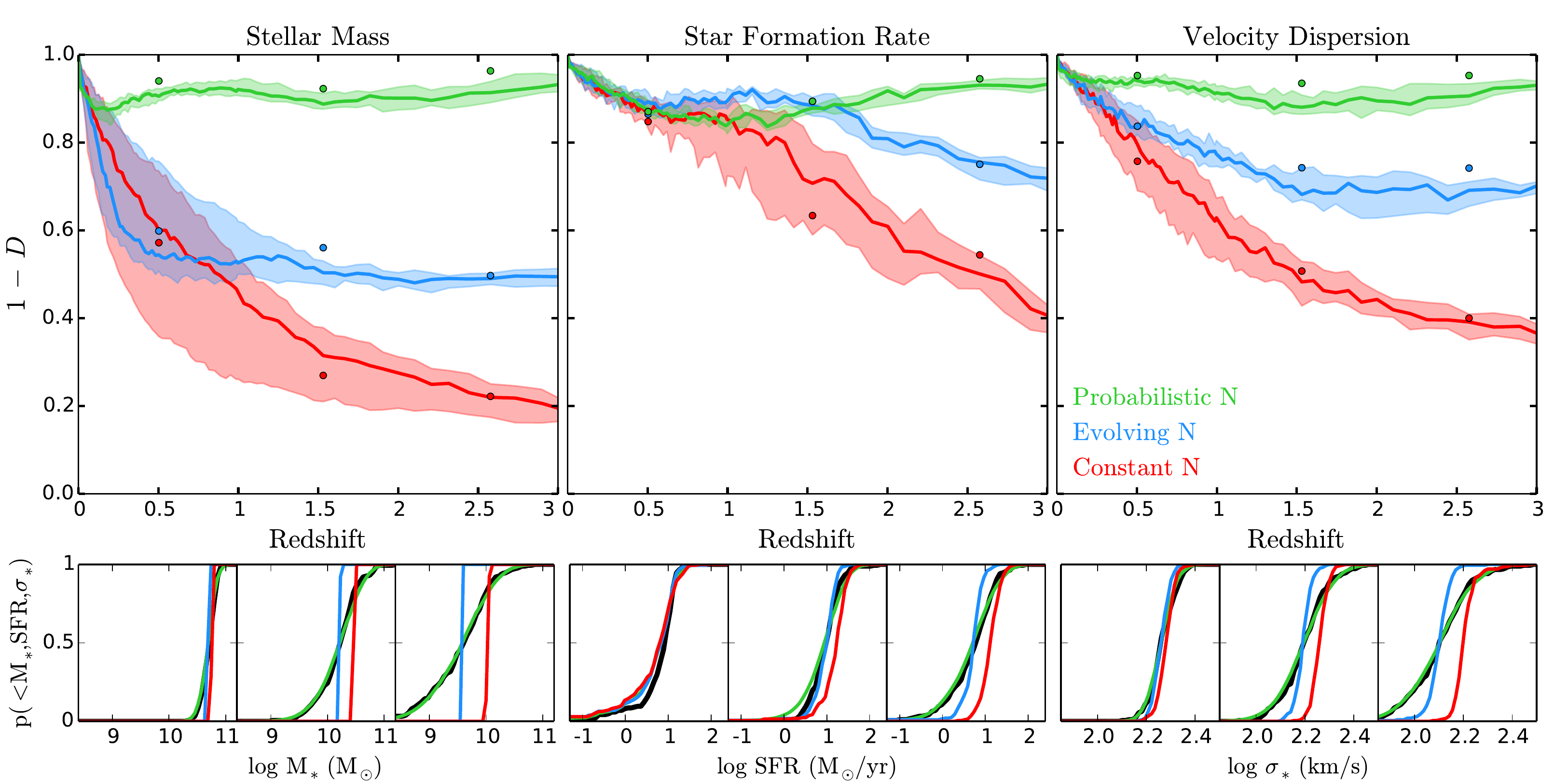}
  \caption{Accuracy of predictions made by different number density methods for the distribution of physical properties within a progenitor population.  We trace samples of 200 galaxies in the Illustris simulation with initial number densities at $z=0$ ranging from $\log N_0 =$ -2 to -4 back to their progenitors at higher redshift, and compare the true distributions of their physical properties against the predicted distributions using the K-S statistic $D$ (the maximum difference between the cumulative distributions in the predicted and true progenitor samples).  Solid lines in the top panels show the median $1-D$ across samples of different $N_0$, and shaded areas the 25th-75th percentile regions, for predictions made under the assumption of a constant (red), evolving (blue), or probabilistic (green) number density.  The bottom panels depict the cumulative distributions at $z$ = 0.5, 1.5, and 2.5 for a single sample with $\log N_0 = -3$, with a black line showing the true distribution of a given property and the colored lines showing the predicted distributions.  Small circles in the upper panels mark the $1-D$ corresponding to these distributions.}
  \label{fig:recoveryfraction}
\end{figure*}

The constant number density method (red) consistently overpredicts the true stellar mass for the Illustris dataset (see the top left panel).  Both the constant and evolving number density methods produce a very narrow range in stellar mass.  The evolving number density method (blue) captures the median mass of the true progenitors.  However, the best agreement for both the median and spread of the true population is achieved with the probabilistic number density method (green).

The following two panels in Figure \ref{fig:predictfromsample} show predictions for the star formation rate and stellar velocity dispersion derived from the stellar-mass-selected progenitor galaxy populations.  Velocity dispersions for the simulated galaxies are calculated by adding in quadrature the standard deviation of the x, y, and z-velocities of all star particles within the central 5 kpc and dividing by $\sqrt{3}$.  Velocity dispersions for the observational sample are derived from the fundamental plane as described in Section \ref{ssec:sample}.  The lines are faint below log $\sigma_*$/(km/s) = 1.95 as a reminder that these values are extrapolations.  The same patterns hold for these quantities as for stellar mass: a systematic offset in the constant number density prediction, the evolving number density method capturing the median but not the spread, and the probabilistic method successfully capturing both.  (Interestingly, the probabilistically-predicted distribution in SFR is slightly too wide at $z\sim 1.5$ -- see Section \ref{ssec:limits} for a discussion.)  Unlike stellar mass, the SFR and stellar velocity dispersion are not included in our fitting procedure, and so they act as an independent verification of the utility of this method.  

When sampling from a population which follows observational relations (bottom), the median predicted stellar masses at $z=3$ differ by a factor of four between the constant number density method and the evolving number density methods.  These different predictions for stellar mass result in predictions for star formation rate and stellar velocity dispersions which differ by approximately the same factor.  

The median values predicted by the evolving and probabilistic methods evolve closely together.  If a median value is all that is desired, the evolving number density method may therefore suffice.  However, the evolution of the population median is not representative of the paths taken by individual galaxies -- the evolution {\it of} the median is not equivalent to evolution {\it along} the median.  As before, the distributions predicted by the probabilistic method are wider.  These different distribution shapes will affect any quantity which is an integral over the entire population.  For example, calculating the fraction of mass growth from in-situ star formation requires a comparison between the evolution of the mean (linear) mass and star formation rate.  Each of these quantities requires knowledge of the full distribution in order to compute the mean.  Thus, though the evolving and probabilistic methods predict similar medians, they could predict quite different results for this or any other integrated quantity. 

\section{Discussion}
\label{sec:discussion}

\subsection{Quantifying prediction accuracy}
\label{ssec:accuracy}

In the previous section, we demonstrated the application of three different number density methods to predict the progenitors of Milky-Way-mass galaxies.  We argued that allowing the median number density to evolve with redshift produced better results than assuming it to be constant, and allowing it to have a distribution rather than limiting it to a fixed bin better still.  But what exactly do we mean by ``better"?  What are we trying to achieve with these methods, and what is the metric by which we should measure success?

If we simply want to find the median progenitor or descendant, the fixed-bin evolving number density performs well -- but we should not confuse ``median" with ``typical."  The properties of the median galaxy are a poor representation of the properties of the majority of progenitors, so if we want to talk about entire populations and the full range of evolutionary pathways available to individual galaxies, we need to consider entire distributions.

One possible metric for success (when making comparisons using theoretical models where the true progenitors are known) is the fraction of individual progenitor galaxies which are accurately identified by each method (the ``recovery fraction").  With the simulation data, we can easily calculate the recovery fraction by comparing the unique identities of galaxies' progenitors and descendants at other redshifts.  For the fixed-bin (i.e. constant and evolving number density) methods, the recovery fraction is straightforwardly computed by counting the number of galaxies which appear in both the true and predicted samples.  For the probabilistic method, the recovery fraction is the sum of the progenitor probabilities assigned to each of the true progenitors.  

When we calculate recovery fraction, we find that all three number density methods perform similarly.  In fact, the variation in recovery fraction between the different predictive methods is dwarfed by the variation produced by changes in the initial number density or size of the galaxy sample.  For a sample size of 200 galaxies, the median recovery fraction across samples with initial number densities $\log N_0 =$ -2 to -4 falls below 20\% by $z=0.5$ and asymptotically approaches $\sim5$\%.  However, the recovery fraction for individual samples ranges from 80\% to 5\% depending on $N_0$.  Recovery fraction is similarly sensitive to sample size, as was shown by \citet{Leja2013}.  See the Appendix for a further discussion of the effects of changes in initial number density and sample size.  The recovery fraction is hence a poorly-defined and unstable quantity.  Moreover, it has no real meaning when applied to observations, since individual galaxies at different redshifts are not actually progenitors/descendants of one another as they are in simulations.

Recovering individual galaxies may therefore not be the most appropriate goal.  We propose instead that accurate recovery of progenitor properties is a more useful metric for comparing linking methods.

In Figure \ref{fig:recoveryfraction}, we show how well each predictive method reproduces the physical properties of the progenitor population, as quantified by the Kolmogorov-Smirnov statistic, $D$.  Here instead of comparing the identities of individual galaxies between populations, we are comparing the  stellar masses, star formation rates, and velocity dispersions.  In the leftmost panel, we take as the first sample the stellar masses of the galaxies in the true progenitor population, and as a second sample the stellar masses of the galaxies in the predicted populations.  (For the distribution method, we assemble a sample by randomly drawing galaxies from the global population according to progenitor probability.)  We then compare these two samples using the Kolmogorov-Smirnov test, which is a measure of the likelihood of the two samples having been drawn from the same underlying probability distribution.  The quantity on the y-axis is 1 - $D$, where $D$ is the maximum distance between the cumulative probability distributions of the two samples.  Thus a higher y-value means that the predicted sample more closely follows the distribution of the true sample.  We repeat this procedure for 20 samples with different initial number densities and show the median and 25th-75th percentile region of $D$ among those samples.  The smaller panels below show the cumulative distributions from which $D$ is derived for the sample with $\log N_0 = -3$ at $z =$ 0.5, 1.5, and 2.5.

The results for stellar mass reflect the patterns we saw in Section \ref{sec:MW}.  The predictions behave as we expect, given that we performed the number density fits for the purpose of recovering the stellar mass distributions: the constant number density method does a poor job of recovering the stellar masses, the evolving number density slightly better, and the probabilistic number density best of all.  

The remaining two panels show the same exercise for progenitors' star formation rates and velocity dispersions.  These properties were not included in the number density fits, but since both of them correlate with stellar mass their predictions will change as well.  Here, though not quite as strong, we find the same general trends as for stellar mass.  Thus, a better prediction for the stellar mass distribution directly results in a better prediction for other quantities.  

\subsection{Limitations and possible refinements}
\label{ssec:limits}

We have demonstrated in the preceding analysis that the use of an evolving probability distribution for number density can improve predictions of progenitor properties, and can be performed robustly with observational data.  The probabilistic method does no better than other number density methods at predicting the progenitors {\it themselves}, however, and so the results are driven by general trends in the global galaxy population.

The dilution of the signal from the true progenitors means that this method contains an implicit assumption: that the properties of the progenitors at a given stellar mass are representative of (or follow the same distributions as) the global population at that mass.  It is easy to think of scenarios where we could expect this assumption to be violated.  As an example, consider a population of galaxies with stellar mass $10^{11}$ \Msun~at $z=2$.  We might expect that any of their $z=0$ descendants which still lie in that same mass range should be systematically smaller (because they are older) than the general $10^{11}$ \Msun~population \citep[see e.g.,][]{Wellons2016}.  The method that we have presented here would not take this into account, and would overpredict the sizes of those descendants.

Another example of this effect is visible in the top middle panel of Figure \ref{fig:predictfromsample} which shows the predictions for star formation rate in the Illustris simulation.  The distribution of SFR predicted by the probabilistic method (in green) is noticeably wider than the distribution of SFR for the true progenitors (in grey) at $z \sim 1$, despite the very similar stellar mass distributions (left panel).  This overprediction of the width is driven by the assumption that the SFRs follow the main-sequence distribution at a given mass.  All of the progenitors must reach the same stellar mass by $z=0$, however, so the lower-mass progenitors should have systematically high SFRs and the higher-mass progenitors should have systematically low SFRs.  Thus the SFR distribution of the true progenitors is narrower than predicted.  The same effect appears in the middle panel of Figure \ref{fig:recoveryfraction} where the evolving $N$ method (accidentally) performs better than the probabilistic $N$ method until the SFRs decouple from the $z=0$ stellar mass around $z=1.5$.  This type of behavior cannot be captured using any of the number density methods we have discussed so far.

It may be possible to further refine predictions by introducing priors when selecting progenitors from a given stellar mass bin.  For the first example, the selection might be improved by looking for the oldest galaxies.  In a similar way, one might assume that the central density or velocity dispersion should undergo minimal evolution, and use it to inform progenitor selection.  One could also look for a progenitor sample which has the same clustering properties as the original sample, or use separate number density tracks for star-forming and quiescent galaxies as suggested by \citet{Clauwens2016}.  These refinements, however, would introduce increasingly model-dependent biases.  The balance to be struck between generality and increased accuracy will depend on the particular application.

\section{Conclusion}
\label{sec:conclusion}

In this paper we have presented an improvement to methods which use cumulative comoving number density to connect galaxy populations across time.  We address the evolution and spread in galaxy number densities by taking a probabilistic, rather than deterministic, approach when selecting galaxy progenitors and descendants.  We draw upon theoretical results for the evolution of the distribution of galaxy number densities within a population to define a probability that a given galaxy at $z_f$ is a progenitor/descendant of a galaxy at another redshift $z_0$.  

From these progenitor probabilities, we can then construct distributions of galaxy properties (e.g. stellar mass, star formation rate) within the progenitor/descendant population by integrating over the entire sample using the progenitor probabilities as weights.  This approach allows us to make more accurate predictions about these properties because we take the full distribution of number densities (i.e. stellar masses) into account.  We have shown that we are able to more accurately recover the distributions not only of stellar mass, but also of secondary inferred quantities such as star formation rate and velocity dispersion.

This probabilistic method is equally applicable to data from simulations and observations, provided the appropriate cumulative mass function is used to convert between number density and stellar mass.  We provide a tool for performing this type of analysis on observational data at \url{https://github.com/sawellons/NDpredict}.

\section*{Acknowledgments}

SW is supported by the National Science Foundation Graduate Research Fellowship under grant number DGE1144152.  PT acknowledges support from NASA ATP Grant NNX14AH35G.  We thank Rachel Bezanson for helpful discussions and guidance regarding the observational aspects of the paper, and Lars Hernquist for his feedback on the draft.

\appendix

\section{Variation in recovery fractions due to changes in initial number density and bin size}
\label{app:softening}

In this Appendix, we examine the sensitivity of the recovery fractions and K-S statistic $D$ discussed in Section \ref{ssec:accuracy} to variations in the initial number density and size of the sample.  The recovery fraction compares the \textit{identities} of the galaxies in a predicted progenitor population against the true progenitors as traced through the simulation, and is simply calculated as the fraction of true progenitors which are accurately predicted by a given number density method.  The K-S statistic $D$ is used to compare the \textit{properties} of the predicted and true samples, e.g. stellar mass, star formation rate, or velocity dispersion, and is calculated as the maximum difference between the cumulative distributions of that property from the two samples.  Each of these quantities will be affected by the initial number density of the sample as well as the sample size.

\begin{figure*}
  \centering
  \includegraphics[width=2.1\columnwidth]{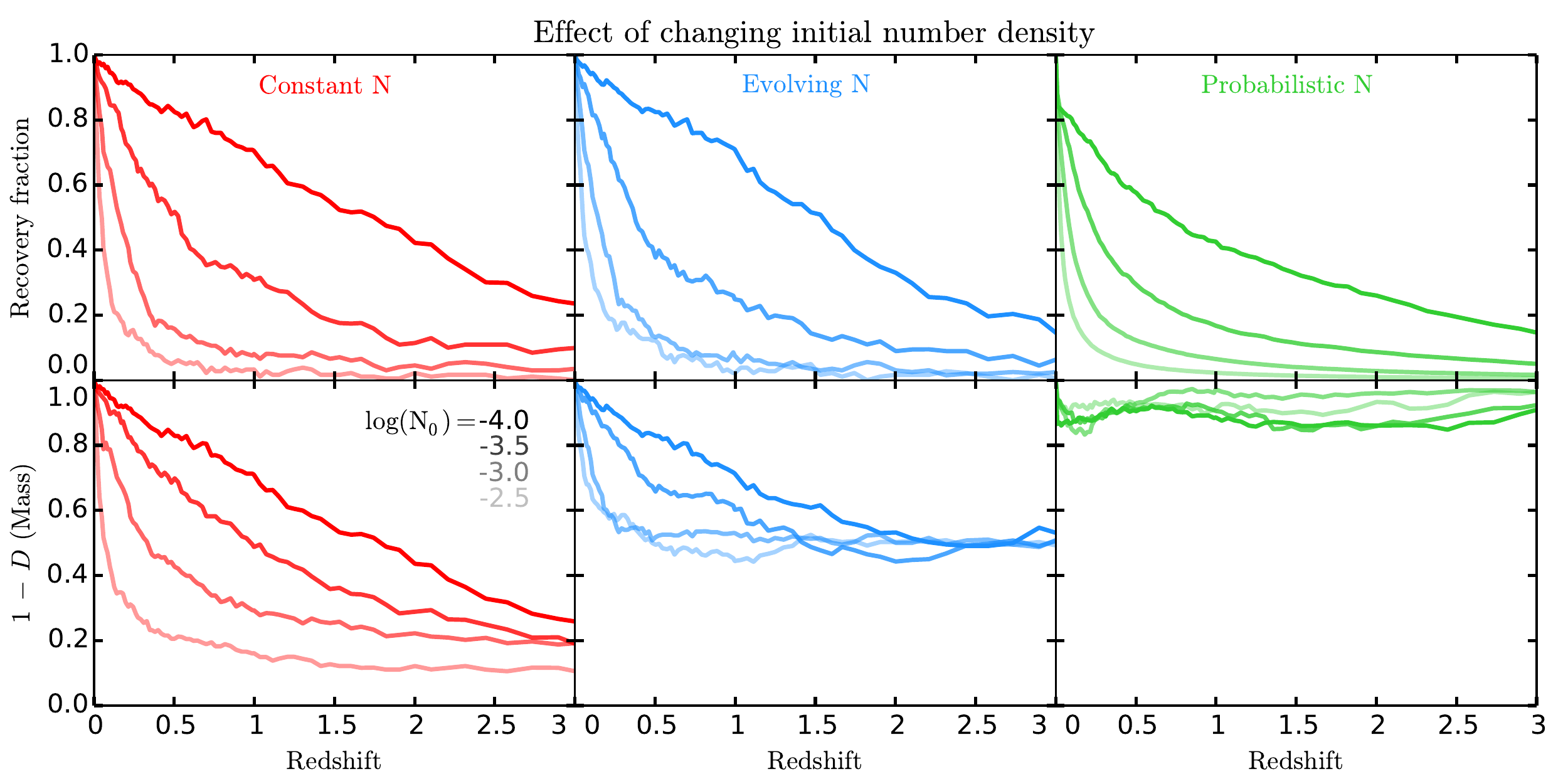}
  \caption{Evolution of recovery fraction and the K-S statistic $D$ for stellar mass for galaxy samples initially selected at $z=0$ and traced/predicted at higher redshift.  The samples each include 200 galaxies, but vary in initial redshift from $\log(N_0)$ = -4 (the darkest lines in each panel) to $\log(N_0)$ = -3.5, -3, and -2.5 (the lightest lines).  The quantities are shown for the predictions made by the constant number density, evolving number density, and probabilitstic number density methods in the left, middle, and right columns respectively.  In all cases, recovery fraction is very sensitive to the initial number density of the sample, while the asymptotic behavior of $D$ remains relatively consistent with the exception of the constant $N$ method.}
  \label{fig:rf_initN}
\end{figure*}

The shaded regions in Figure \ref{fig:recoveryfraction} show the scale of the variation between samples at different initial number densities.  In Figure \ref{fig:rf_initN}, we show the evolution in recovery fraction and $1-D$ (for stellar mass) for four specific samples of 200 galaxies at initial number densities $\log(N_0)$ = -4, -3.5, -3, and -2.5.  Each column shows predictions using a different number density method.  In the case where the probabilistic number density assumption is employed, we can see that the recovery fraction starts out at lower values near $z=0$ than the other methods.  This is a result of the assumption that the true number densities instantly follow a lognormal distribution, which is not immediately true as they relax away from their initial delta function.  Regardless of the prediction method, the recovery fraction evolves strongly with initial number density.  

The K-S statistic $D$, on the other hand, is less sensitive to this perturbation.  The clear exception here is the case when a constant number density assumption is employed, but this is now a meaningful variation -- the mass function is steepest at the high-mass (low-number density) end, so a misprediction of the number density will not impact the stellar mass prediction as much.  For the other two methods which accurately track the bulk motion, the asymptotic behavior of $D$ does not change very much with $N_0$.  In particular, $D$ for the evolving number density method will always asymptotally approach 0.5 by definition, since the predicted cumulative mass distribution is essentially a step function at the median of the true distribution.  

\begin{figure*}
  \centering
  \includegraphics[width=2.1\columnwidth]{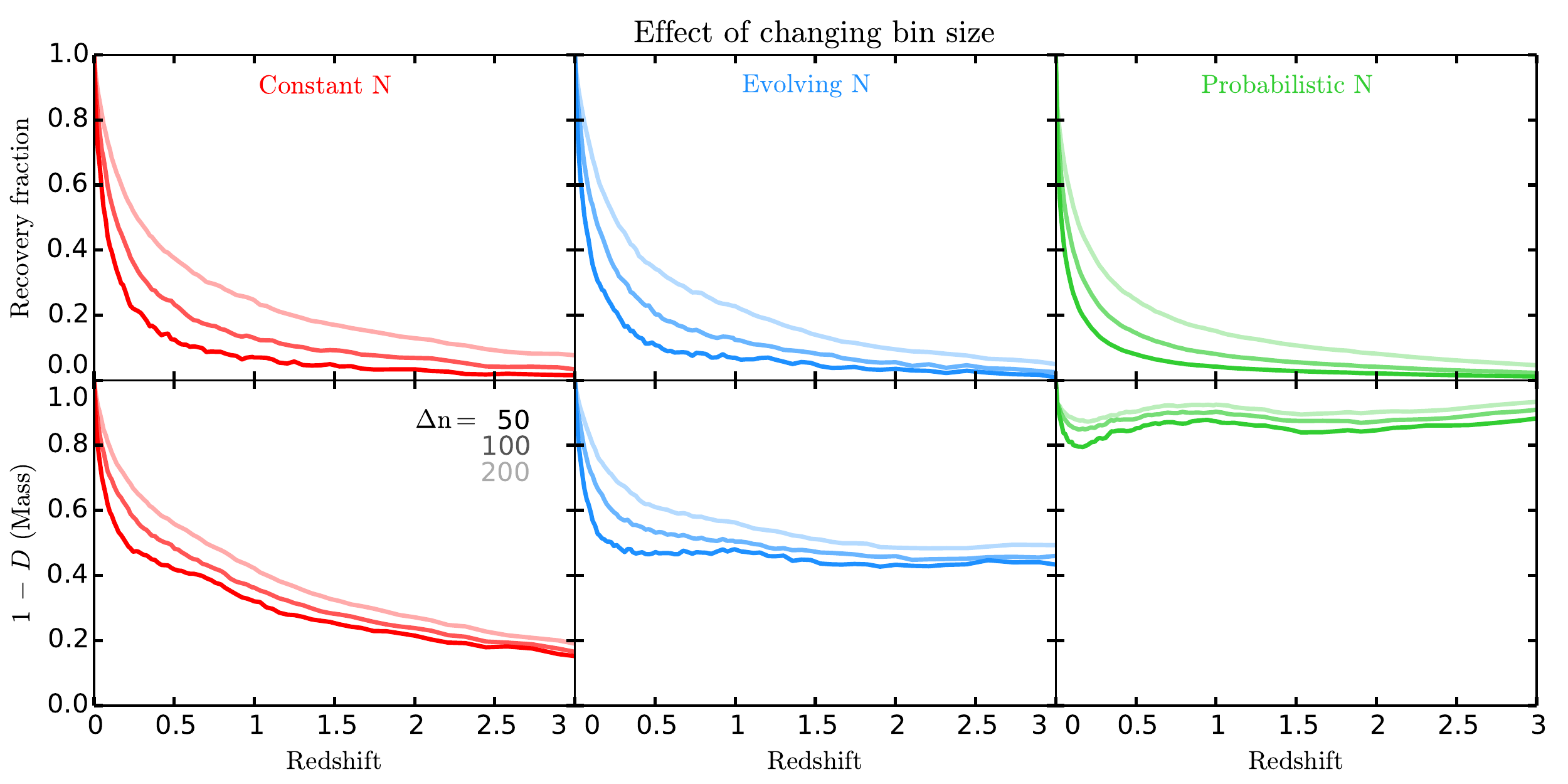}
  \caption{Evolution of recovery fraction and the K-S statistic $D$ for stellar mass for galaxy samples initially selected at $z=0$ and traced/predicted at higher redshift.  The samples vary in size from 50 galaxies (the darkest lines) to 200 galaxies (the faintest lines), and the line shown is the average quantity for samples ranging from initial number densities of $\log(N_0)$ = -4 to -2.5.  The quantities are shown for the predictions made by the constant number density, evolving number density, and probabilisttic number density methods in the left, middle, and right columns respectively.  In all cases, recovery fraction is sensitive to the size of the sample, as is $D$ to a somewhat lesser degree.}
  \label{fig:rf_binsize}
\end{figure*}

The recovery fraction is also sensitive to changes in the bin/sample size, as has already been documented for descendant galaxies by \citet{Leja2013}.  In Figure \ref{fig:rf_binsize}, we show the recovery fraction and $1-D$ for galaxy samples of size $n$ = 50 (the darkest lines), 100, and 200 (the faintest lines).  As in Figure \ref{fig:recoveryfraction}, the lines shown are averages over samples with $\log(N_0)$ = -4 to -2.5.  In general, the larger the sample, the less noise and the more reliably one can capture some of its behavior.  Both the recovery fraction and (to a somewhat lesser degree) $1-D$ are sensitive to this effect, although the relative performance of the number density methods remains the same.

The relative insensitivity of $D$ to these perturbations make it a better metric by which to compare predicted and true progenitor/descendant populations than recovery fraction, particularly in light of the fact that the recovery fraction is meaningless as applied to observational data.

\label{lastpage}

\end{document}